\documentclass[twocolumn,secnumarabic,amssymb, nobibnotes,prd,superscriptaddress]{revtex4}

\usepackage[utf8]{inputenc}
\usepackage{graphics,graphicx}
\usepackage{physics}
\usepackage{amsmath,amssymb,amsfonts,latexsym,cancel}
\usepackage{multirow}

\usepackage{bm}
\begin{document}

\title{Fluid pulsation modes and tidal deformability of anisotropic strange stars in light of the GW$170817$ event}

\author{Jos\'e D. V. Arba\~nil}
\email{jose.arbanil@upn.pe}
\affiliation{Departamento de Ciencias, Universidad Privada del Norte, Avenida el Sol 461 San Juan de Lurigancho, 15434 Lima, Peru}
\affiliation{Facultad de Ciencias F\'isicas, Universidad Nacional Mayor de San Marcos, Avenida Venezuela s/n Cercado de Lima, 15081 Lima, Peru}
\author{Cesar V. Flores}
\email{cesarovfsky@gmail.com}
\affiliation{Departamento de F\'isica-CCET, Universidade Federal do Maranh\~ao, Campus Universitário do Bacanga, S\~ao Lu\'is CEP 65080-805, MA, Brazil}
\affiliation{Centro de Ci\^encias Exatas, Naturais e Tecnol\'ogicas, Universidade Estadual da Regi\~ao Tocantina do Maranh\~ao, Imperatriz CEP 65901-480, MA, Brazil}
\author{C\'esar H. Lenzi}
\email{chlenzi@ita.br}
\affiliation{Departamento de F\'isica, Instituto Tecnol\'ogico de Aeron\'autica, Centro T\'ecnico Aeroespacial, 12228-900 S\~ao Jos\'e dos Campos, S\~ao Paulo, Brazil}
\author{Juan M. Z. Pretel}
 \email{juanzarate@cbpf.br}
 \affiliation{
 Centro Brasileiro de Pesquisas F{\'i}sicas, Rua Dr. Xavier Sigaud, 150 URCA, Rio de Janeiro CEP 22290-180, RJ, Brazil
}
%
%

\date{\today}

\begin{abstract}
The effects of the anisotropy on the fluid pulsation modes adopting the so-called  Cowling approximation and tidal deformability of strange quark stars are investigated by using the numerical integration of the hydrostatic equilibrium, nonradial oscillations, and tidal deformability equations, being these equations modified from their standard form to include the anisotropic effects. The fluid matter inside the compact stars is described by the MIT bag model equation of state. For the anisotropy profile, we consider a local anisotropy that is both regular at the center and null at the star's surface. We find that the effect of the anisotropy is reflected in the fluid pulsation modes and tidal deformability. Finally, we analyze the correlation between the tidal deformability of the GW$170817$ event with the anisotropy.
\end{abstract}

\maketitle


\section{\label{sec:level1} Introduction}

\smallskip

With all the recent detections of gravitational signals coming from the merger of binary systems reported by LIGO-Virgo Collaboration (LVC) \cite{abbott2016,abbott2016_2,abbott2017_4,abbott_2018a_tidal,abbott2017,abbott2017_1,abbott2017_2,abbott2017_3,abbott2016_3,abbott2017_5}, we can say that we live at the beginning of a new golden age in general relativity, the age of gravitational wave astronomy.  In this sense, it is essential to invest our best efforts in order to study new quantitative, qualitative, and even exotic physical characteristics that could be present in future multi-messenger observations.

\smallskip

Among those phenomena, it is well known that compact stars can be present as components of binary systems. Then the behavior of a compact star prior to, during, and after the merger can not be ignored. For example, when the binary system is very close, the tidal interaction plays an important role \cite{abbott_2018a_tidal}, it could be a natural route to obtain information about the equation-of-state (EOS) within the signals emitted during a merger of two compact stars. 

\smallskip

The theoretical methods used to investigate the oscillation frequencies of stars is asteroseismology. This theory is a powerful tool that gives us a firm path in the search of traces of physics inside compact stars \cite{pulsating1b,pulsating2c,pulsating2a, pulsating2b}. In this way, the oscillation frequencies of such stars, namely, $f$- and $p$-modes \cite{stellarpul1,stellarpul2}, would give us information about the composition and internal structure of such spherical objects (check, e.g., \cite{pulsating2, pulsating3, pulsating4, pulsating5, flores2017} and their references).

\smallskip

An important aspect to be analyzed in the study of compact objects is the tidal deformability \cite{flanagan,hinderer_2008,damour2009,postnikov2010,poisson}. As previously mentioned, this parameter gives us information about the equation of state hidden in the signals emitted by compact stars. Moreover, nowadays, using dimensionless tidal deformability, we can place some limits on the theory using the observational data of the event GW$170817$.

\smallskip

In this regard, we investigate the non-radial oscillation modes and tidal deformability of anisotropic strange quark stars. As reported, theoretical evidence indicates that anisotropies can emerge in highly dense media, for instance, such as that appear in phase transitions \cite{paper3}, pion condensed phase \cite{paper4}, a solid or superfluid nucleus \cite{ruderman_1972, heiselberg_2000}, or in the presence of superfluid type 3A \cite{paper5}. Since establishing a connection between the internal composition of the compact star and the results reported by observation is the purpose of many works, the tidal deformability results found in this work are analyzed in light of the deformability parameter obtained from the event GW$170817$, see Ref. \cite{abbott_2019}.

\smallskip

This article is structured as follows: In Sec. \ref{section2}, we present the Einstein field equation, energy-momentum tensor, stellar structure equations, non-radial oscillation equations, and tidal deformability equations and their boundary conditions. In Sec. \ref{section3} we show the numerical method employed, the EOS, the anisotropic profile, and the scaling solution for non-radial oscillation equations and for tidal deformability equations. Moreover, in this section, we plot the change of oscillation frequency and tidal deformability against anisotropy. Finally, in Sec. \ref{section4} we conclude. Throughout the entire work, in order to simplify our equations and also for numerical reasons, we will employ the units $G=1=c$.

\section{General relativistic formulation}\label{section2}
\subsection{Einstein field equation}\label{subsection1}

We start by writing the Einstein field equation in the presence of an anisotropic fluid:
\begin{equation}\label{EFeq}
    G^{\mu}_{\varphi}=8\pi T^{\mu}_{\varphi} ,
\end{equation}
where the Greek indexes $\mu$, $\varphi$, etc.~go from $0$ to $3$; $G^{\mu}_{\varphi}$ is the Einstein tensor, and $T^{\mu}_{\varphi}$ represents the energy-momentum tensor which is given by
\begin{equation}\label{tem}
    T^{\mu}_{\varphi}=(\rho+p_t)u^{\mu}u_{\varphi}+p_t g^{\mu}_{\varphi}-\sigma k^{\mu}k^{\nu}g_{\nu\varphi},
\end{equation}
with $\rho$, $p_t$, and $\sigma=p_t-p_r$ being respectively the energy density, tangential pressure, and anisotropic pressure parameter; where $p_r$ is the radial pressure. Besides, $u_{\varphi}$ is the four-velocity of the fluid, $k_{\varphi}$ denotes the radial unit vector, and $g_{\mu\varphi}$ stands the metric tensor. These $4$-vectors must satisfy the following conditions:
\begin{eqnarray}
    k_{\varphi}k^{\varphi}=1,\hspace{0.4cm}
    u_{\varphi}k^{\varphi}=0,\hspace{0.4cm}
    u_{\varphi}u^{\varphi}=-1.
\end{eqnarray}

\subsection{Static background equations}\label{subsection2}

The unperturbed spherically symmetric line element, in Schwarzschild-like coordinates, is expressed in the form
\begin{equation}\label{metric}
    ds^2=-e^{2\Phi}dt^2+e^{2\Psi}dr^2+r^2(d\theta^2+\sin^2\theta d\phi^2),
\end{equation}
where the metric functions $\Phi=\Phi(r)$ and $\Psi=\Psi(r)$ depend on the radial coordinate $r$ alone.

Considering the spacetime metric \eqref{metric} and the potential metric 
\begin{equation}\label{g00}
    e^{-2\Psi}=\left(1-\frac{2m}{r}\right),
\end{equation}
the non-null components of the field equations \eqref{EFeq} can be placed into the form
\begin{eqnarray}
&&m'=4\pi\rho r^2,\label{mass_conservation} \label{eq_mass}\\
&&p_r'=-(p_r+\rho)\left[4\pi rp_r +\frac{m}{r^2}\right]e^{2\Psi}+\frac{2\sigma}{r},\label{tov_equation}\\
&&\Phi'=-\frac{p_r'}{\rho+p_r}+\frac{2\sigma}{r(\rho+p_r)}.\label{dg11dr}
\end{eqnarray}
The function $m(r)$ is the mass enclosed within the sphere radius $r$. Eqs.~\eqref{mass_conservation} and \eqref{tov_equation} represent respectively the mass conservation and the hydrostatic equilibrium equation \cite{tolman,oppievolkoff} modified from the original form to include the anisotropic factor \cite{paper2}. This set of equations is known as the stellar structure equations. The prime $(\,'\,)$ over the functions stands for the derivation concerning $r$.

To obtain the stellar equilibrium configurations, we integrate Eqs.~\eqref{eq_mass}-\eqref{dg11dr} from the origin up to the radial coordinate where the radial pressure vanishes. In other words, the solution starts at the center of the star ($r=0$), where
\begin{align}
    m(0) &= 0,  &  \Psi(0) &=0,  &  \Phi(0)& =\Phi_c,  &  \rho(0) &=\rho_{c},
\end{align}
and the stellar surface ($r=R$) is determined by
\begin{eqnarray}
p_r(R)=0.
\end{eqnarray}
In addition, at $r=R$, the interior spacetime metric connects smoothly with the Schwarzschild vacuum exterior solution, so that
\begin{equation}\label{surface_condition}
e^{2\Phi}=e^{-2\Psi}=1-\frac{2M}{R},
\end{equation}
with $M$ standing for the total mass of the star.

\subsection{Non-radial oscillations equations within the Cowling approximation}\label{subsection2}

In non-radial oscillations of compact stars, the Cowling formalism \cite{mcdermott1983, lindblom1990} is often used to calculate the oscillation frequencies (see, e.g., Refs.~\cite{vasquez2014, sotani2011}), since this method provides a good accuracy of the oscillation frequencies obtained from the numerical approach of complete general relativity. In fact, in typical stellar models, between both methods, a discrepancy of less than $20\%$ and $10\%$ for the $f$ and $p_1$-modes are respectively found \cite{yoshida1997}. This good precision justifies its use to study, for example, the fluid pulsation mode of neutron stars in the presence of slow \cite{stavridis2007} and rapid rotation rate \cite{boutloukos2007}, crust elasticity \cite{samuelsson2007}, internal anisotropy \cite{doneva2012}, and $d$-dimensions \cite{arbanil2020}. 

To investigate pulsation modes of anisotropic strange stars, the metric functions are kept fixed in the Cowling approximation, i.e., $\delta g_{\mu\nu}=0$ \cite{sotani2011}. In addition, the equations describing the fluid pulsation are obtained by perturbing the conservation equation of the energy-momentum tensor \eqref{tem}. We hence obtain $\delta\left(\nabla_{\mu}T^{\mu\nu}\right)=0$. Projecting this relation both along the four-velocity $u_{\nu}$ and orthogonally by employing the operator ${\cal P}_{\mu}^{\nu} = \delta_{\mu}^{\nu}+u^{\nu}u_{\mu}$, we get respectively:
\begin{align}
&u^{\nu}\nabla_{\nu}\delta\rho+\nabla_{\nu}\left(\left[(\rho+p_t)\delta^{\nu}_{\mu}-\sigma k^{\nu}k_{\mu}\right]\delta u^{\mu}\right)\nonumber\\
&\quad +(\rho+p_t)a_{\nu}\delta u^{\nu}-\nabla_{\nu}u_{\mu} \delta\left(\sigma k^{\nu}k^{\mu}\right)=0,\label{nro_eq1} \\
&\left(\delta\rho+\delta p_t\right)a_{\mu}+\left(\rho+p_t\right)u^{\nu}\left(\nabla_{\nu}\delta u_{\mu}-\nabla_{\mu}\delta u_{\nu}\right)\nonumber\\
&\quad +\nabla_{\mu}\delta p_t+u_{\mu}u^{\nu}\nabla_{\nu}\delta p_t-{\cal P}_{\mu}^{\nu}\nabla_{\alpha}\delta\left(\sigma k^{\alpha}k_{\nu}\right)=0,\label{nro_eq2}
\end{align}
with $a_{\mu}=u^{\nu}\nabla_{\nu}u_{\mu}$ being the four-acceleration.

Taking into account the Lagrangian fluid vector components in the form
\begin{equation}
\xi^{i}=\left(e^{-\Psi}W,-V\partial_{\theta},-\frac{V}{\sin^{2}\theta}\partial_{\phi}\right)\frac{Y_{\ell m}}{r^2},
\end{equation}
with $W=W(t,r)$ and $V=V(t,r)$ being functions of the coordinates $t$ and $r$, and $Y_{\ell m}=Y_{\ell m}(\theta,\phi)$ are the spherical harmonics. In such a way, the perturbation of the four-fluid through the Lagrangian perturbation vector $\delta u^{\mu}=\left(0,\delta u^r,\delta u^{\theta},\delta u^{\phi}\right)$ can be expressed as
\begin{equation}
\delta u^{\mu}=\left(0,e^{-\Psi}\partial_{t}W,-\partial_{t}V\partial_{\theta},-\frac{\partial_{t}V}{\sin^{2}\theta}\partial_{\phi}\right)\frac{Y_{\ell m}e^{-\Phi}}{r^2}.
\end{equation}

Considering $u^{\mu}=\left(e^{-\Phi},0,0,0\right)$, $k^{\mu}=\left(0,e^{-\Psi},0,0\right)$, $\sigma=\sigma(p_r,\mu)$, $W(t,r)=W(r)e^{i\omega t}$, and $V(t,r)=V(r)e^{i\omega t}$ in Eqs.~\eqref{nro_eq1} and \eqref{nro_eq2}, we arrive at the following system of equations
\begin{align}
    W' =&\ \frac{d\rho}{dp_r}\left[ \left(1+{\cal X}\right)\left(1+\frac{\partial\sigma}{\partial p_r}\right)^{-1}\frac{\omega^2r^2V}{e^{2\Phi-\Psi}}+\Phi'W \right] \nonumber  \\
    &- {\cal X}\left[ \left(1+\frac{d\rho}{dp_r}\right)\frac{2W}{r}+\ell(\ell+1)e^{\Psi}V \right]  \nonumber \\
    &- \ell(\ell+1)e^{\Psi}V ,  \label{nro_eq_w}  \\
    V' =&\ V\left[ -\frac{\sigma'}{\rho+p_r+\sigma}-\left(\frac{d\rho}{dp_r}+1\right)\left(\Phi'+\frac{2}{r}\right)\frac{{\cal X}}{1+{\cal X}} \right.  \nonumber \\
    &\left.+ \frac{2}{r}\frac{\partial\sigma}{\partial p_r}+\left(1+\frac{\partial\sigma}{\partial p_r}\right)^{-1}\left(\frac{\partial^2\sigma}{\partial p_r^2}p_r'+\frac{\partial^2\sigma}{\partial p_r\partial\mu}\mu'\right)\right]  \nonumber \\
    &+ 2V\Phi'-\left(1+\frac{\partial\sigma}{\partial p_r}\right)\frac{1}{1+{\cal X}}\frac{e^{\Psi}W}{r^2} , \label{nro_eq_v} 
\end{align}
where we have defined ${\cal X}= \sigma/(\rho+p_r)$ and, following Ref.~\cite{doneva2012}, we consider $\delta\sigma= (\partial\sigma/\partial p_r)\delta p_r$, with $\delta\mu=0$ $\left(\mu=2m/r\right)$, and $\omega$ representing the oscillation eigenfrequency. These two differential equations are reduced to those found in \cite{sotani2011} taking $\sigma=0$.

To solve Eqs.~\eqref{nro_eq_w} and \eqref{nro_eq_v} from the center $(r=0)$ toward the stellar surface $(r=R)$, we need to impose boundary conditions. Thus, at $r=0$, we consider that the functions $W$ and $V$ assume the respective forms
\begin{equation}
W=Cr^{\ell+1},\quad\quad V=-C\frac{r^{\ell}}{\ell},
\end{equation}
with $C$ representing a dimensionless constant. Moreover, at $r=R$ (where $p_r=0$) is found that
\begin{equation}\label{surface_condition_NRO}
\left[1+{\cal X}\right]\frac{\omega^2V}{e^{2\Phi}}+\left[1+\frac{\partial\sigma}{\partial p_r}\right]\left[\frac{r\Phi'}{2}-{\cal X}\right]\frac{2W}{e^{\Psi}r^3}=0.
\end{equation}
Hereafter, to compare our results with those shown in the literature, e.g., Refs.~\cite{doneva2012,aquino2022}, we restrict our results to the quadripolar modes ($\ell=2$).

\subsection{Tidal deformability}\label{subsection2}

The study of the theory of tidal Love numbers is frequent within the context of binary compact star systems. In this scheme, the gravitational effects caused by one star can outcome in the deformation of its companion. Such deformation, produced by an external field, can be measured through the tidal deformability parameter $\lambda$. This parameter can be expressed as follows
\begin{equation}
\lambda=-\frac{Q_{ij}}{\epsilon_{ij}}, 
\end{equation}
with $Q_{ij}$ and $\epsilon_{ij}$ representing the quadrupole moment and an external tidal field, see Refs.~\cite{hinderer_2008, damour2009, hinderer_2010}. The relation that directly connects the tidal deformability parameter and the quadripolar Love number $k_2$ is given by
\begin{equation}
    k_2=\frac{3}{2}\lambda R^{-5}.
\end{equation}

Furthermore, the dimensionless tidal deformability $\Lambda$, as a function of the Love number $k_2$, follows from the following relation
\begin{equation}\label{tidal_deformability}
    \Lambda=\frac{2k_2}{3C^5},
\end{equation}
where $C=M/R$ represents the compactness parameter. $k_2$ can also be written in terms of $C$. Thus, we have
\begin{align}
{k_2} =&\ \frac{8C^5}{5}(1-2C)^2[2+C(y_R-1)-y_R] \nonumber \\
&\times\{2C [6-3y_R+3C(5y_R-8)]+4C^3[13-11y_R \nonumber \\
&+C(3y_R-2)+2C^2(1+y_R)]+3(1-2C^2) \nonumber  \\
& \times [2-y_R+2(y_R-1)]\ln(1-2C)\}^{-1},
\end{align}
with the function $y_R=y(r=R)$. In addition, the function $y(r)$ satisfies the Riccati differential equation
\begin{equation}\label{riccati_eq}
y'r+y^2+y\left(C_0r-1\right)+C_1r^2=0,
\end{equation}
where
\begin{align}
    C_0 =&\ \frac{2m}{r^2}e^{2\Psi}+4\pi e^{2\Psi}\left(p_r-\rho\right)r+\frac{2}{r}, \label{C_0_y}  \\
    C_1 =&\ 4\pi e^{2\Psi}\left[4\rho+4p_r+4p_t+\frac{p_r+\rho}{Ac_s^2}\left(c_s^2+1\right)\right] \nonumber \\
    &-\frac{6}{r^2}e^{2\Psi}-4\Phi'^2, \label{C_1_y}
\end{align}
with $c_s^2=\frac{dp_r}{d\rho}$ and $A=\frac{dp_t}{dp_r}$. Comparing Eqs.~\eqref{C_0_y} and \eqref{C_1_y} with the forms presented in Refs.~\cite{biswas2019, Rahmansyah2020}, we see that our $C_0$ and $C_1$ are in agreement and contradiction, respectively, with those presented in these two works of literature. Note that the first-order differential equation \eqref{riccati_eq} is derived from the second-order differential equation for the function $H$ in the quadripolar case $(\ell=2)$, Eq.~\eqref{EDO_H}, by using $y= rH'/H$. Moreover, if we consider $p_t=p_r$ (i.e., $A=1$), Eq.~\eqref{riccati_eq} is reduced to the isotropic case (see \cite{postnikov2010}).

In particular, for strange quark stars---where the energy density at the star's surface is finite and non-null---it is required a correction term in the calculation of 
$y_R$. Thus, due to this energy discontinuity $y_R$ stays \cite{damour2009,wang2019,li2020,zhou2018,lourenco2021}
\begin{equation}
y_R \longrightarrow y_R-\frac{4\pi R^3 \rho_s}{M},
\end{equation}
where $\rho_s$ represents the energy density difference between the internal and external regions.

\section{Results}\label{section3}

\subsection{Numerical method}

To investigate the influence of anisotropy in the oscillation spectrum and tidal deformability of strange stars --once defined the EOS and the anisotropy profile-- the stellar structure equations \eqref{g00}{-\eqref{tov_equation}, the non-radial oscillation equations \eqref{nro_eq_w}-\eqref{nro_eq_v}, and the tidal deformability equations \eqref{tidal_deformability}-\eqref{C_1_y} are integrated from the center $(r=0)$ to the star's surface $(r=R)$. 

To study both the fluid perturbation modes and the tidal deformability, we first integrate Eqs. \eqref{g00}-\eqref{dg11dr} from the center to the star's surface through the fourth-order Runge-Kutta method, for different values of $\kappa$ and $\rho_c$. Once determine the parameters $p_r$, $p_t$, $\rho$, $m$, and $\Phi$, Eq. \eqref{dg11dr} is solved by using the shooting method. This integration begins considering a proof value of $\Phi_c$, if after the numerical solution the equality shown in Eq. \eqref{surface_condition} is not attained, $\Phi_c$ is corrected until satisfying this condition. 

The numerical solution of the nonradial oscillations and tidal deformability equations are respectively described below:
\begin{itemize}
    \item The fluid perturbation modes equations, Eqs. \eqref{nro_eq_w}-\eqref{nro_eq_v}, are integrated from the center to the star's surface. It begins by taking into account the correct value of $\Phi_c$ in the stellar structure equations for a particular value of $\kappa$ and $\rho_c$, and the test value of $\omega^2$. If the numerical integration of the equality \eqref{surface_condition_NRO} is not reached, $\omega^2$ is corrected in the next integration until this condition is satisfied.
   
   \item The tidal deformability equations, Eqs. \eqref{tidal_deformability}-\eqref{C_1_y}, are integrated along the radial coordinate $r$ which goes from $0$ to $R$. It starts employing the certain value of $\Phi_c$ in the stellar structure equations for a particular value of $\kappa$ and $\rho_c$.
\end{itemize}

\subsection{Equation of state and anisotropic profile}\label{section_eos}

To depict the strange quark fluid that makes up the compact object, the MIT bag model EOS is employed. This EOS describes a fluid containing only up, down, and strange quarks that have no mass and no interaction, confined by a bag constant ${\cal B}$. For the anisotropic fluid  analyzed, we assume that the radial pressure and the energy density are related by the equality:
\begin{equation}
    p_r=\frac{1}{3}\left(\rho-4\,{\cal B}\right).
\end{equation}
This EOS is widely employed for the theoretical possibility that strange matter can be the ground state of strongly interacting matter and it could appear in compact stars \cite{witten1984}. In \cite{farhi1984} this hypothesis is verified for a bag constant in the range of $57$ and $94\,[\rm MeV/fm^3]$. Following \cite{arbanil_malheiro_2016}, we employ ${\cal B} = 60\,[\rm MeV/fm^3]$.

For the anisotropic pressure profile, inspired in \cite{horvat2011,arbanil_malheiro_2016,arbanil_panotopoulos2022,doneva2012,folomeev2015,silva2015}, we use the quasilocal form $\sigma=\sigma(p_r,\Psi)$. It depends on quantities that remit information on both the state of the fluid and the geometry at a particular interior point of the spacetime. Thus, we consider the anisotropic profile:
\begin{equation}\label{anisotropic_eos}
    \sigma=\kappa p_r\left(1-e^{-2\Psi}\right),
\end{equation}
being $\kappa$ a dimensionless anisotropic constant. The relation \eqref{anisotropic_eos} was used, for instance, to investigate the influence of the anisotropy on the radial oscillation of polytropic stars \cite{horvat2011,{arbanil_panotopoulos2022}} and strange stars \cite{arbanil_malheiro_2016}, nonradial oscillation of neutron stars \cite{doneva2012}, magnetic field structure \cite{folomeev2015}, and slowly rotating neutron stars \cite{silva2015}.

\subsection{Scaling solution of non-radial oscillations equations and tidal deformability equations}

In the literature, it has been reported that when a linear EOS is used to describe the fluid of a star, e.g., the MIT bag model EOS, both the stellar structure and radial oscillation equations admit a scaling law for several star properties \cite{arbanil_malheiro_2016,witten1984,glendenning_book}. This means that if a star's properties are known for a given ${\cal B}$, these properties can be found for another value of ${\cal B'}$. 

For stellar structure, non-radial oscillation, and tidal deformability equations a scaling law can also be used. This can be realized through the following variables:
\begin{eqnarray}
&&{\tilde p}_r=\frac{p_r}{\cal B},\quad{\tilde \rho}=\frac{\rho}{\cal B},\quad{\tilde \sigma}=\frac{\sigma}{\cal B},\quad{\tilde m}={m}\sqrt{\cal B},\nonumber\\
&&{\tilde r}={r}\sqrt{\cal B},\quad{\tilde \omega}=\frac{\omega}{\sqrt{\cal B}},\quad{\tilde W}=\frac{W}{e},\\
&&{\tilde V}=\frac{V}{f},\quad{\tilde C}_0=\frac{C_0}{\sqrt{\cal B}},\quad{\tilde C}_1=\frac{C_1}{{\cal B}},\quad{\tilde y}=y,\nonumber
\end{eqnarray}
where $f=\sqrt{\cal B}e$, and with $f$ and $e$ to be positive and non-null. Considering this scaling law, the stellar structure, non-radial oscillations, and tidal deformability equations keep their original form. Thus, knowing the properties of a star for a certain value of ${\cal B}$, the properties of another star with a different value of ${\cal B'}$ can be determined considering the scale:
\begin{eqnarray}
&&\frac{\rho'_c}{\cal B'}=\frac{\rho_c}{\cal B},\quad M'\sqrt{\cal B'}=M\sqrt{\cal B},\quad R'\sqrt{\cal B'}=R\sqrt{\cal B},\nonumber\\
&&\frac{\omega'}{\sqrt{\cal B'}}=\frac{\omega}{\sqrt{\cal B}},\quad\Lambda'=\Lambda,
\end{eqnarray}
with $\rho_c$ being the central energy density.


\subsection{Oscillation spectrum of the anisotropic strange stars}

\begin{figure*}[ht!]
\centering
\includegraphics[width=8.5cm]{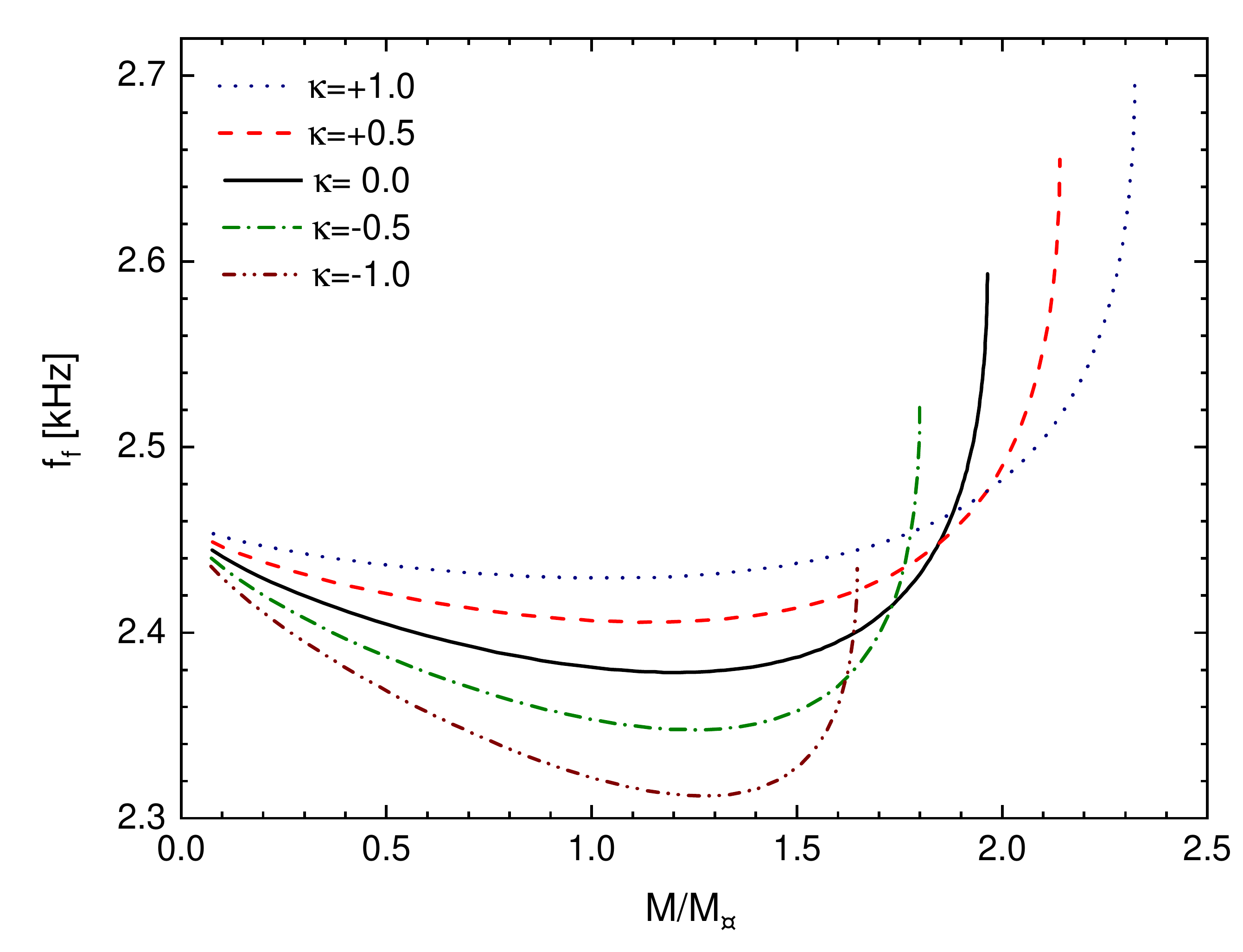} 
\includegraphics[width=8.5cm]{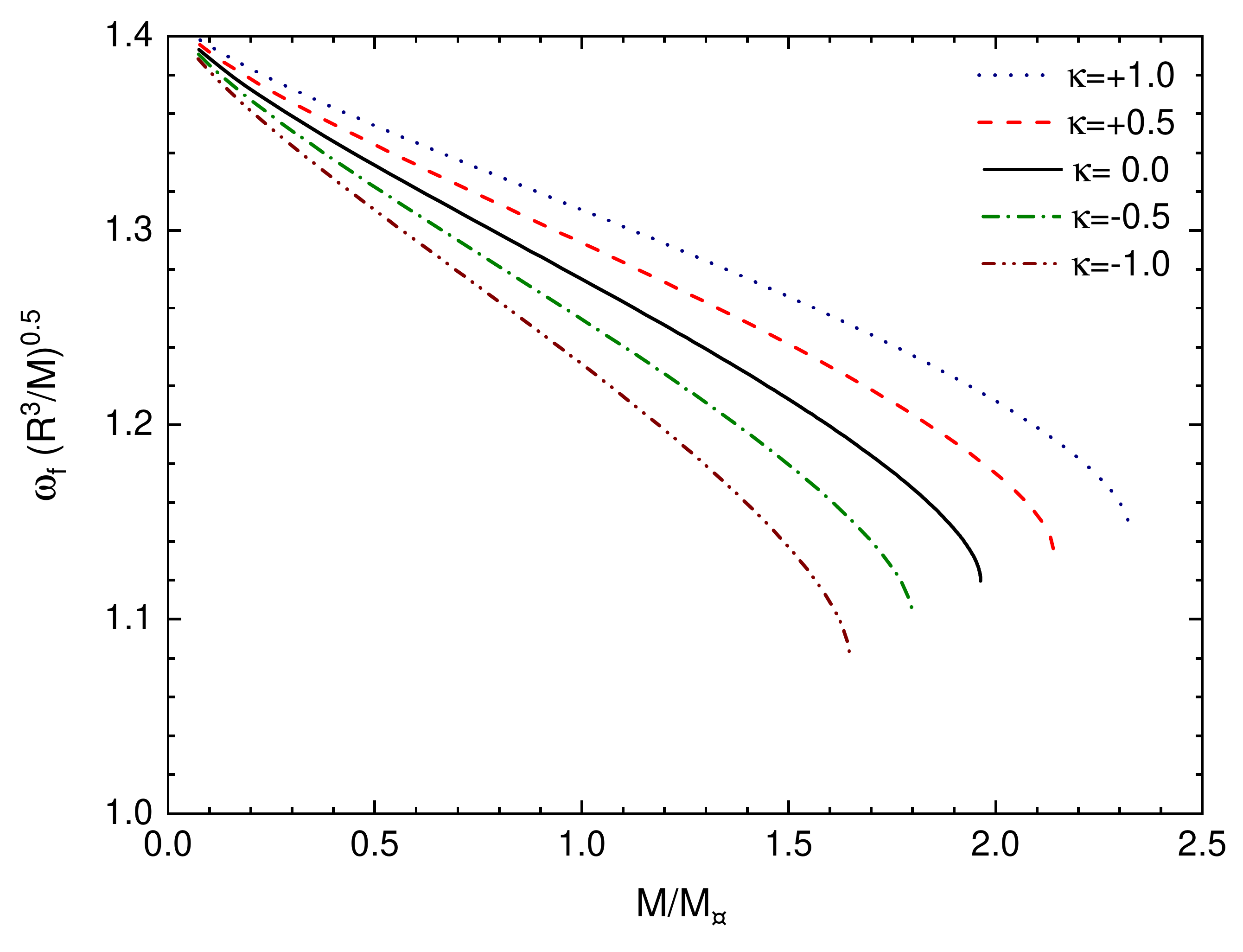}
 \includegraphics[width=8.5cm]{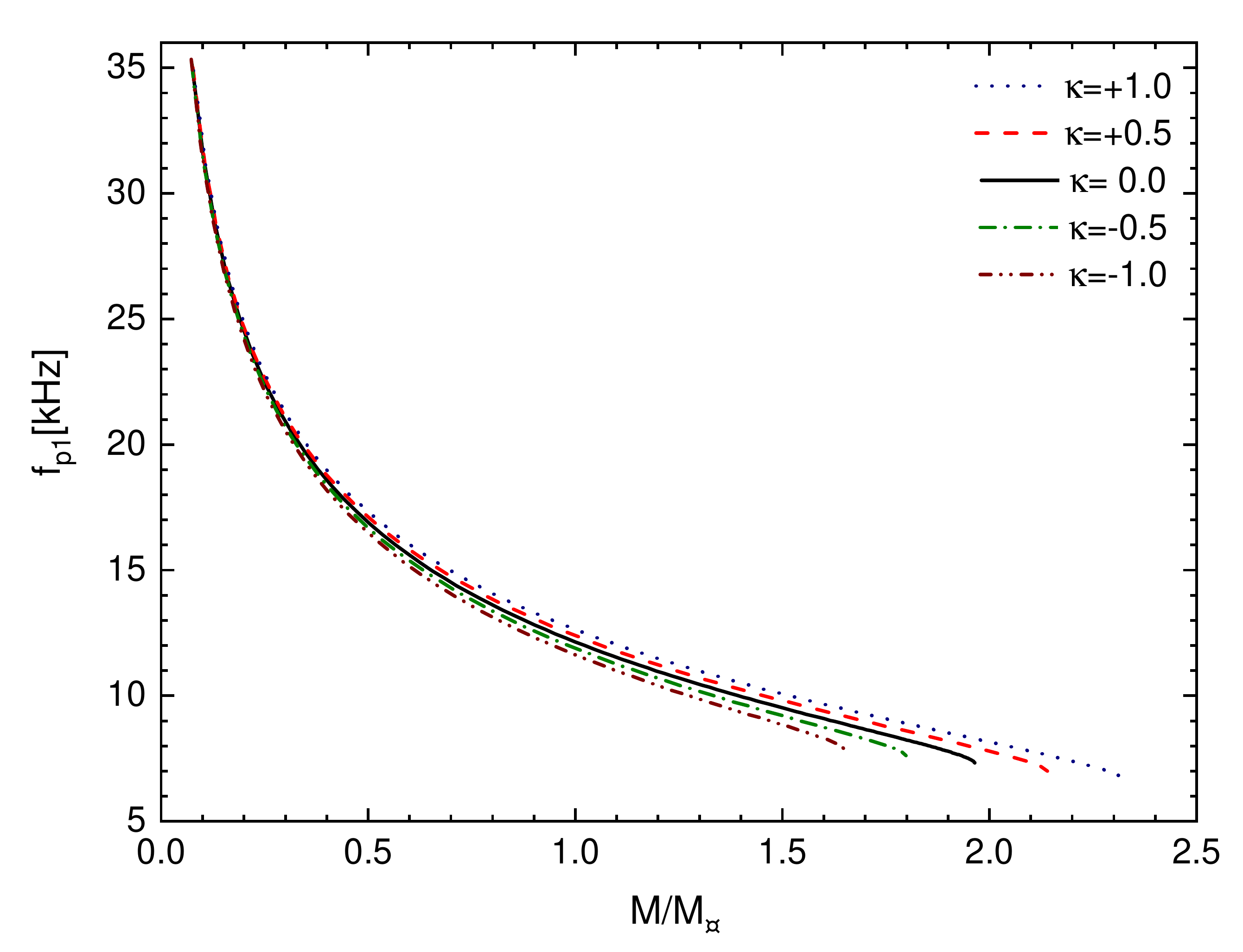} 
\includegraphics[width=8.5cm]{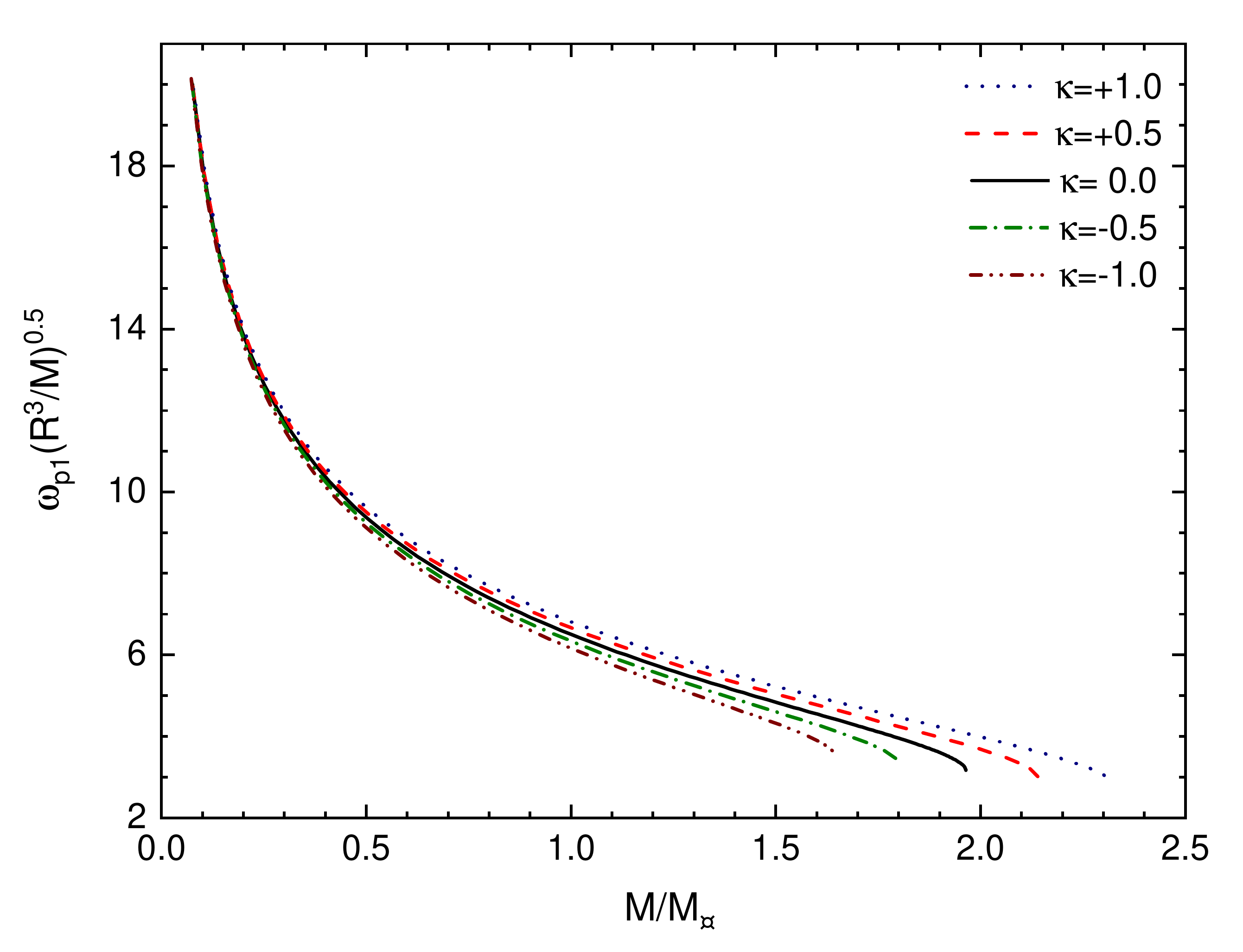} 
\caption{\label{fig1} Upper panels: Oscillation frequency $f_f$ (left side) and normalized frequency $\omega_f$ (right side) as functions of the total gravitational mass for the $f$-mode. Meanwhile, the frequencies corresponding to the $p_1$-mode are shown in the lower plots. We have used five values for the anisotropy parameter $\kappa$, where the isotropic solution is represented by the black curve. It can be observed that the $f$-mode frequencies increase (decrease) because of a positive (negative) anisotropy. Something similar occurs in the case of the $p_1$-mode frequencies, however, the impact of anisotropy is more significant only in the high-mass branch. }
\end{figure*}

The frequency and the eigenfrequency, normalized with the average density $\sqrt{M/R^3}$, versus the total mass $M/M_{\odot}$ are respectively presented on the left and right panels of Fig. \ref{fig1} for five values of $\kappa$. The top and bottom panels show the results for $f$- and $p_1$-modes, respectively. On the left panels, in the $f$-mode frequency case, we note that the curves decrease with the increment of the total mass until attaining a minimum value, after this point the curves turn anti-clockwise to grow with $M/M_{\odot}$. In turn, in the $p_1$-mode frequency case, we obtain that the curves decrease monotonically with the increment of $M/M_{\odot}$. On the right panels, the normalized eigenfrequencies $f$ and in $p_1$ decay monotonically with the growth of $M/M_{\odot}$. Furthermore, from the figures, we can also see that the anisotropy affects the pulsation mode of the fluid. We find that both $f$- and $p_1$-mode change with $\kappa$. For greater $\kappa>0$ (lower $\kappa<0$), stars have a larger (lower) $f_f$, $f_{p_1}$, $\omega_f\left(R^3/M\right)^{0.5}$, and $\omega_{p_1}\left(R^3/M\right)^{0.5}$. This change in frequency is associated with the fact that the radial pressure changes with the anisotropy, see \cite{arbanil_malheiro_2016}.


\subsection{Tidal deformability of the anisotropic strange stars}

The dimensionless tidal deformability as a function of the total mass is shown on the top of Fig. \ref{fig_TD} for different values of $\kappa$. These results are contrasted with the case of $\Lambda_{1.4}=190^{+390}_{-120}$ obtained by LVC \cite{abbott_2018a_tidal}. In all curves, we note that the tidal deformability decreases monotonically with the increment of the total mass. On the other hand, the effects of anisotropy on tidal deformability are also observed. We find that for a larger (lower) value of $\kappa>0$ ($\kappa<0$) greater (lesser) values of $\Lambda$ are derived for the same mass. All these curves are within the range of $\Lambda_{1.4}$ reported by LVC in \cite{abbott_2018a_tidal}. On the bottom of Fig.~\ref{fig_TD}, it is possible to see in more detail the effect of the anisotropic parameter on the $\Lambda_{1.4}$, where the dimensionless tidal deformability undergoes a slight increment (decrement) with increasing (decreasing) of the dimensionless anisotropic constant. 

\begin{figure}
\centering
\includegraphics[width=8.5cm]{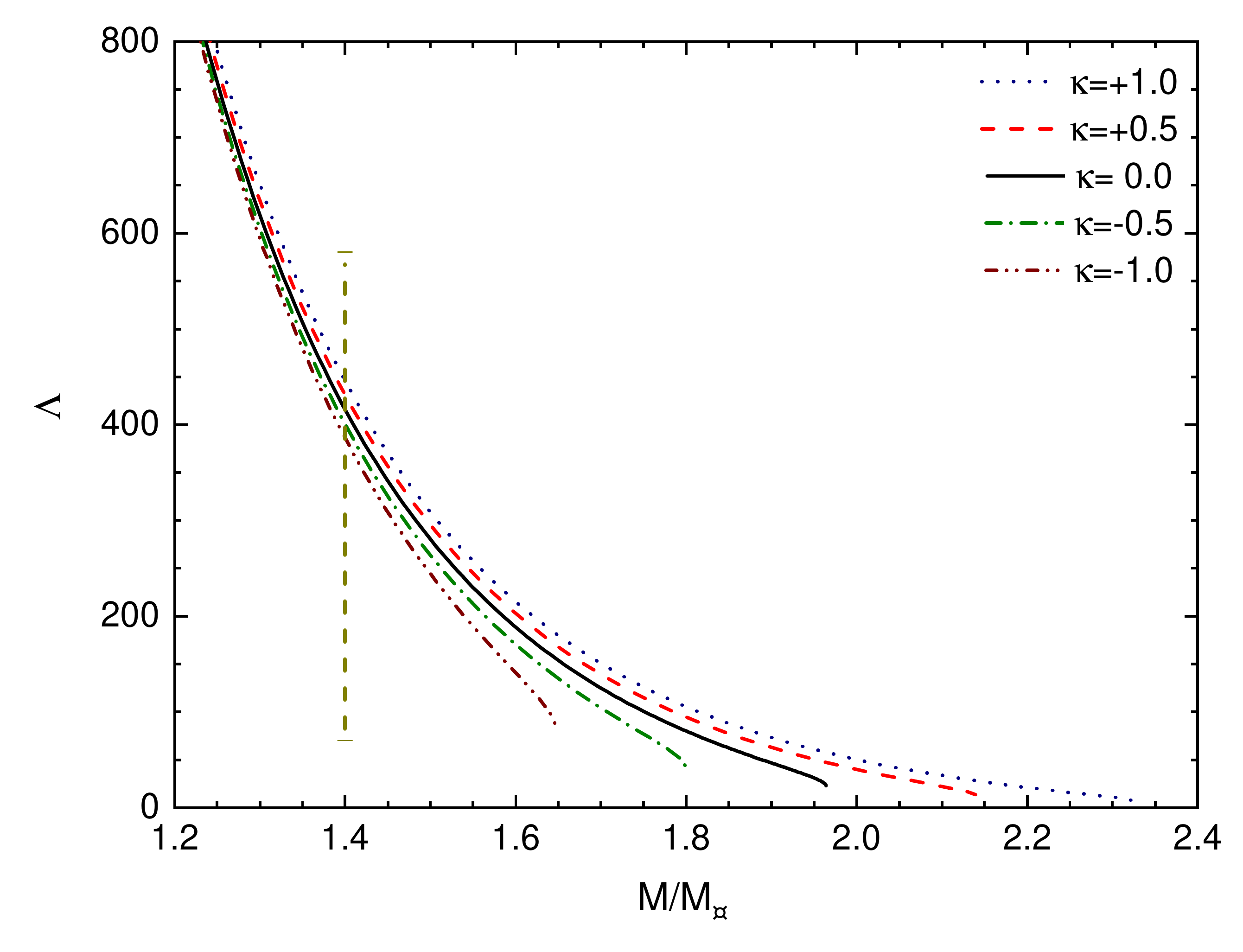} 
\includegraphics[width=8.5cm]{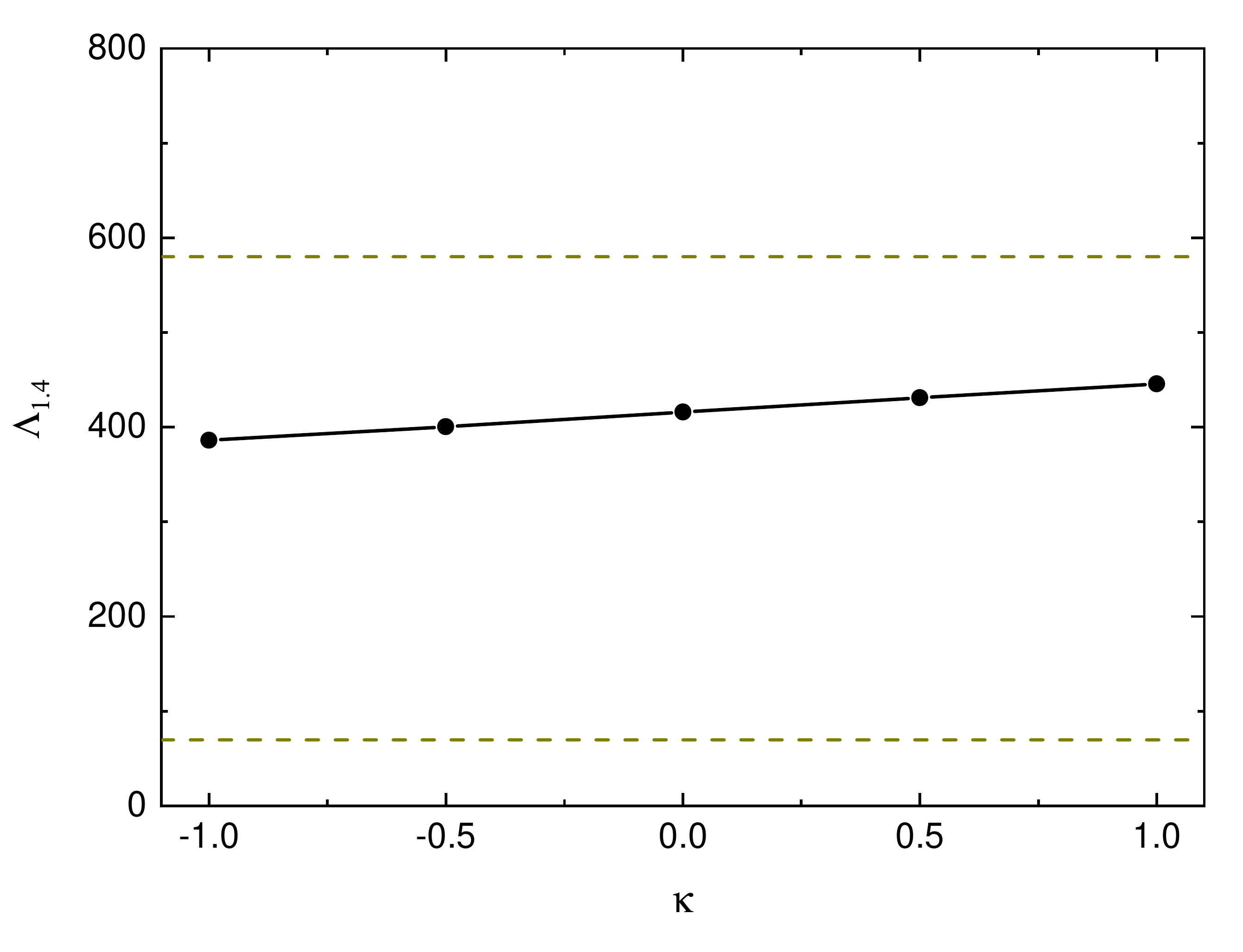} 
\caption{\label{fig_TD} Top: Dimensionless tidal deformability versus the total mass for several values of $\kappa$. Bottom: $\Lambda_{1.4}$ as a function of the dimensionless anisotropic constant $\kappa$. The vertical and horizontal dashed straight lines represent $\Lambda_{1.4}=190^{+390}_{-120}$ reported by LVC in  Ref.~\cite{abbott_2018a_tidal}.}
\end{figure}

On the top and bottom panels of Fig. \ref{fig_ffTD} are respectively shown the oscillation frequency $f_f$ and $f_{p_1}$ against the dimensionless tidal deformability for different values of $\kappa$. These results are contrasted with the $\Lambda_{1.4}=190^{+390}_{-120}$ reported by LVC, check \cite{abbott_2018a_tidal}. In the figure, we note that the $f$-mode ($p_1$-mode) decrease (increase) monotonically with the increment of the dimensionless tidal deformability. In addition, within the interval delimited by the observation, we note that the frequency as a function of the deformability has an almost linear behavior.

\begin{figure}
\centering
\includegraphics[width=8.5cm]{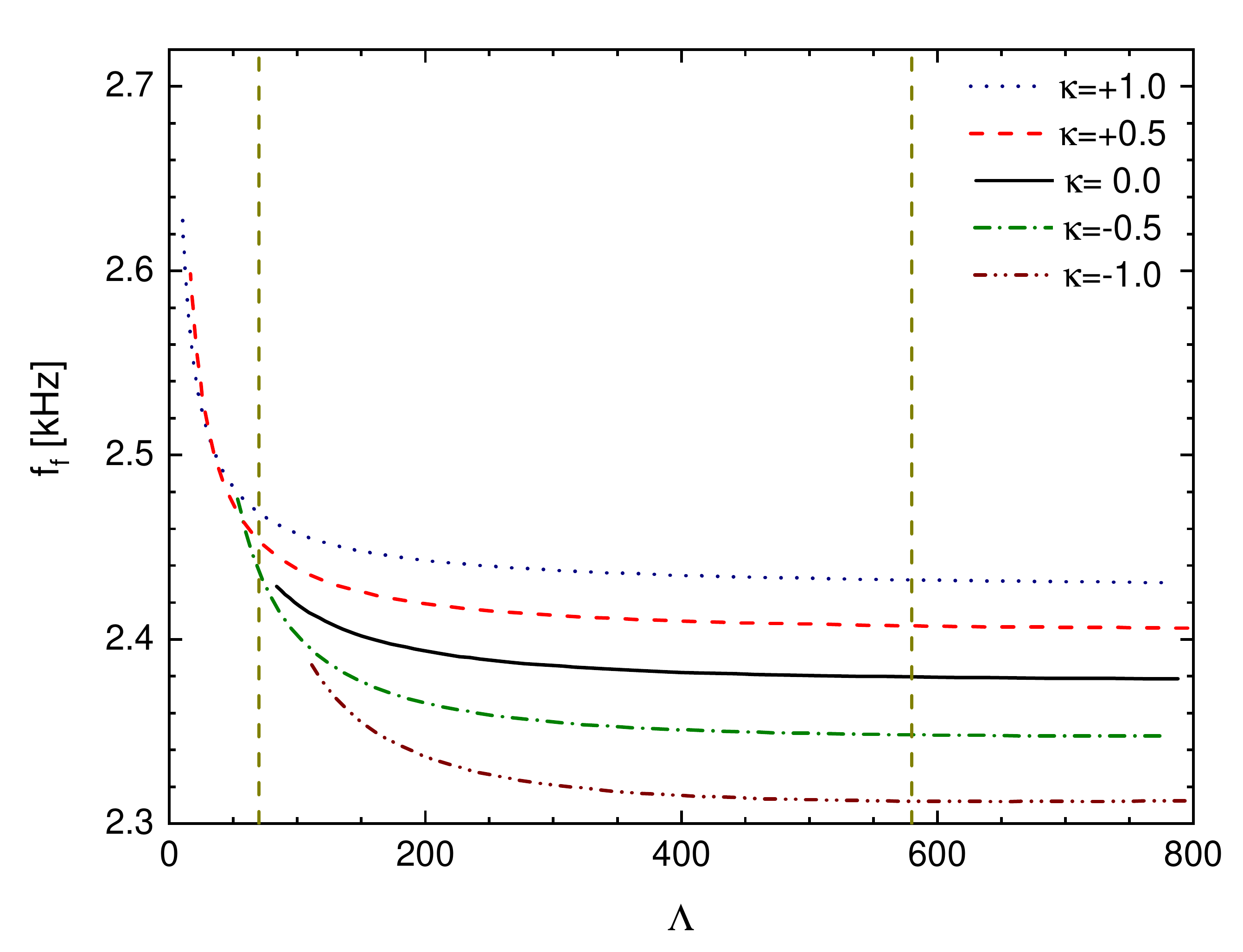} 
\includegraphics[width=8.5cm]{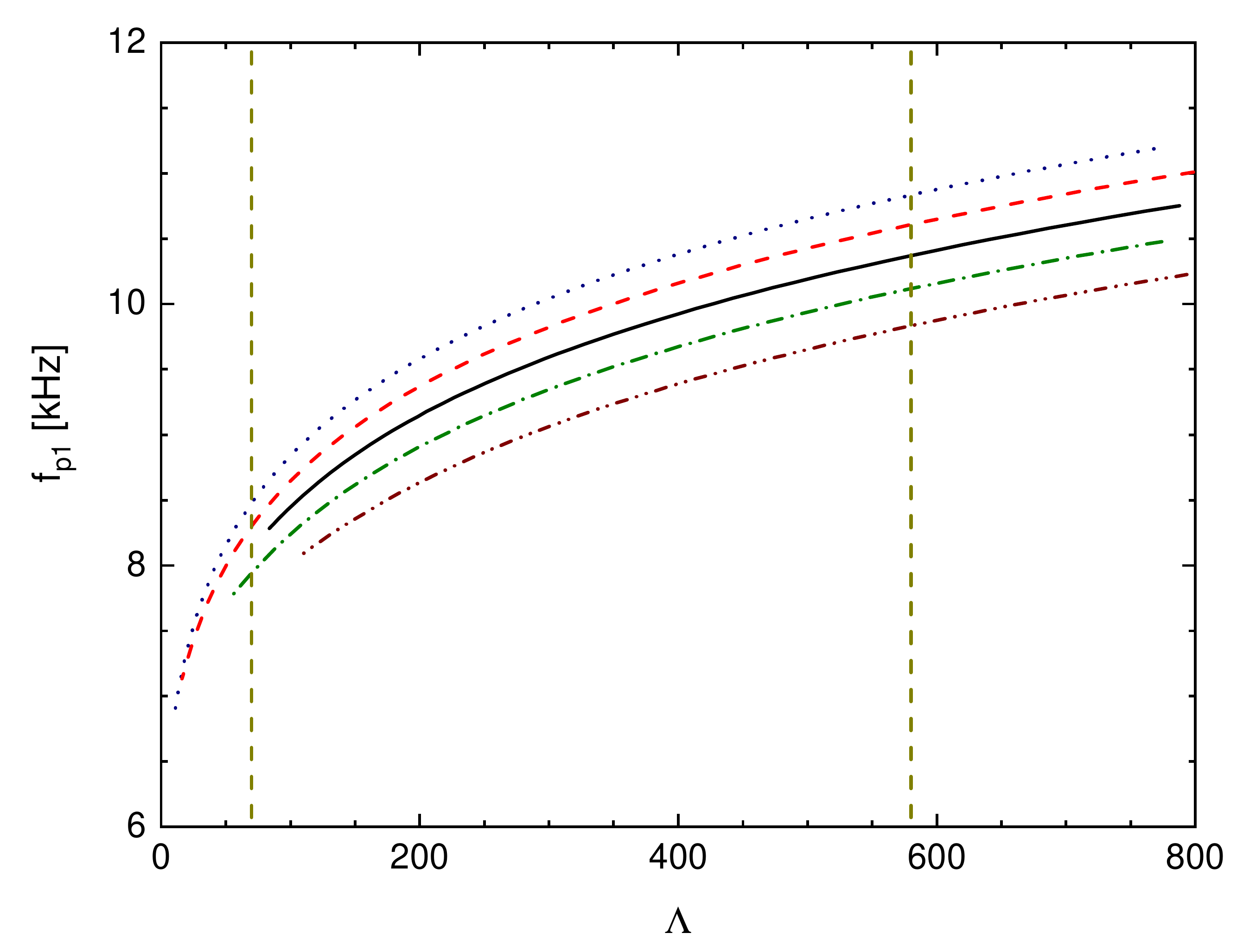} 
\caption{\label{fig_ffTD} Oscillation frequency $f_f$ and $f_{p_1}$ against the tidal deformability for different values of $\kappa$ are plotted on the top and bottom panel. The vertical dashed straight lines mark the tidal deformability $\Lambda_{1.4}=190^{+390}_{-120}$ from the event GW$170817$ estimated in Ref.~\cite{abbott_2018a_tidal}.}
\end{figure}

The obtained data by LVC, allowed authors in \cite{abbott2017_4} to establish some constraints on $\Lambda_1$ and $\Lambda_2$  which are the dimensionless tidal deformability of the binary system, where $\Lambda_1$ is the dimensionless tidal deformability parameter of the star with higher mass in the binary system and $\Lambda_2$ represents the same parameter of the companion star. In Fig.~\ref{fig_2TD}, we plot the diagram $\Lambda_1 \times \Lambda_2$, where the curves  $\Lambda_1-\Lambda_2$ are plotted first chosen a value of $M_1$ and determining $M_2$ via the chirp mass ${\cal M}=1.188\,M_{\odot}$ \cite{abbott2017_4}, defined by
${\cal M}=(M_1\,M_2)^{3/5}/(M_1+M_2)^{1/5}$. Moreover, the values considered for $M_1$ and $M_2$ are within the range $1.36\leq M_1/M_{\odot}\leq 1.60$ and $1.17\leq M_2/M_{\odot}\leq 1.36$, respectively. We also represent the lines of $50\%$ and $90\%$ credibility levels related to the GW$170817$ event established by LVC in the low-spin prior scenario. Either for $\kappa>0$  or $\kappa<0$, we note clearly the influence of the anisotropic parameter on tidal deformability. All curves derived are within the confidence lines taken from Ref. \cite{abbott2017_4}.

\begin{figure}
\centering
\includegraphics[width=8.5cm]{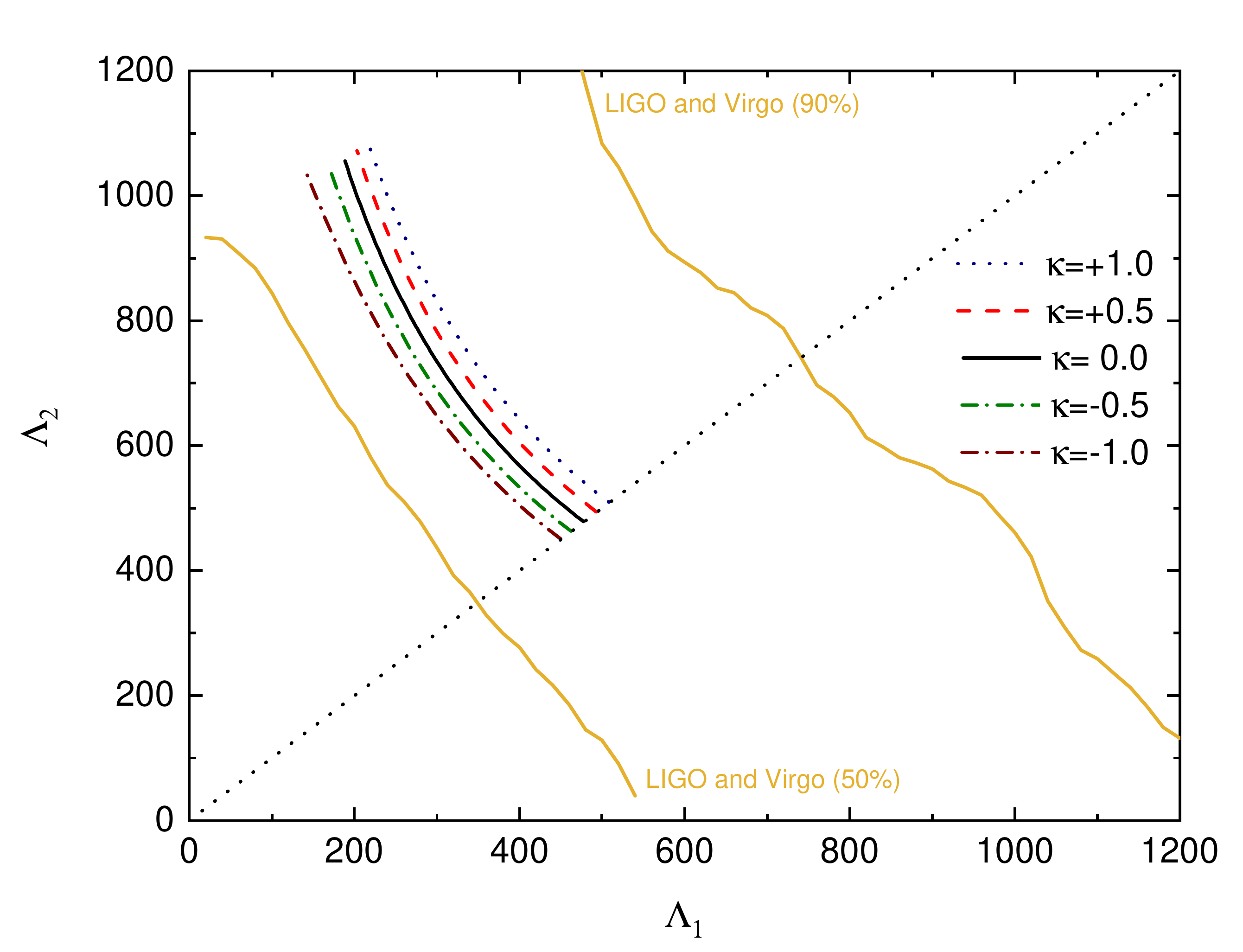} 
\caption{\label{fig_2TD} Dimensionless tidal deformabilities for the components of the GW$170817$ event for different cases of anisotropic parameter $k$. The yellow line represents the LIGO-Virgo confidence curves \cite{abbott2017_4}, and the dotted diagonal line denotes the values that correspond to  $\Lambda_1 = \Lambda_2$.}
\end{figure}

Finally, we study the dimensionless parameter $\tilde\Lambda$, which is measurable through the gravitational waves event of a binary system. $\tilde\Lambda$ is obtained as follows \cite{flanagan}:
\begin{equation}
{\hspace{-0.19cm}}{\tilde\Lambda}=\frac{16}{13}\frac{\left(M_1+12M_2\right)M_1^4\Lambda_1+\left(M_2+12M_1\right)M_2^4\Lambda_2}{\left(M_1+M_2\right)^5}.
\end{equation}
As can be seen, it is calculated using the masses and dimensionless tidal deformability of the stars forming the binary system. Since the masses $M_1(\Lambda_1)$ and $M_2(\Lambda_2)$ are established into a particular interval in agreement with the GW$170817$ event, it is evident that each value of $\kappa$ will produce a range for $\tilde\Lambda$. Thus, in Fig. \ref{fig_TD_tilde} shows $\tilde\Lambda$ against $\kappa$ for anisotropic strange stars. We contrast these results with the constraint on the combined dimensionless tidal deformability reported by LVC, i.e., $\tilde{\Lambda}= 300^{+420}_{-230}$, review Ref. \cite{abbott_2019}. Note that all the obtained values of $\tilde{\Lambda}$ are within the observational intervals.

\begin{figure}
\centering
\includegraphics[width=8.5cm]{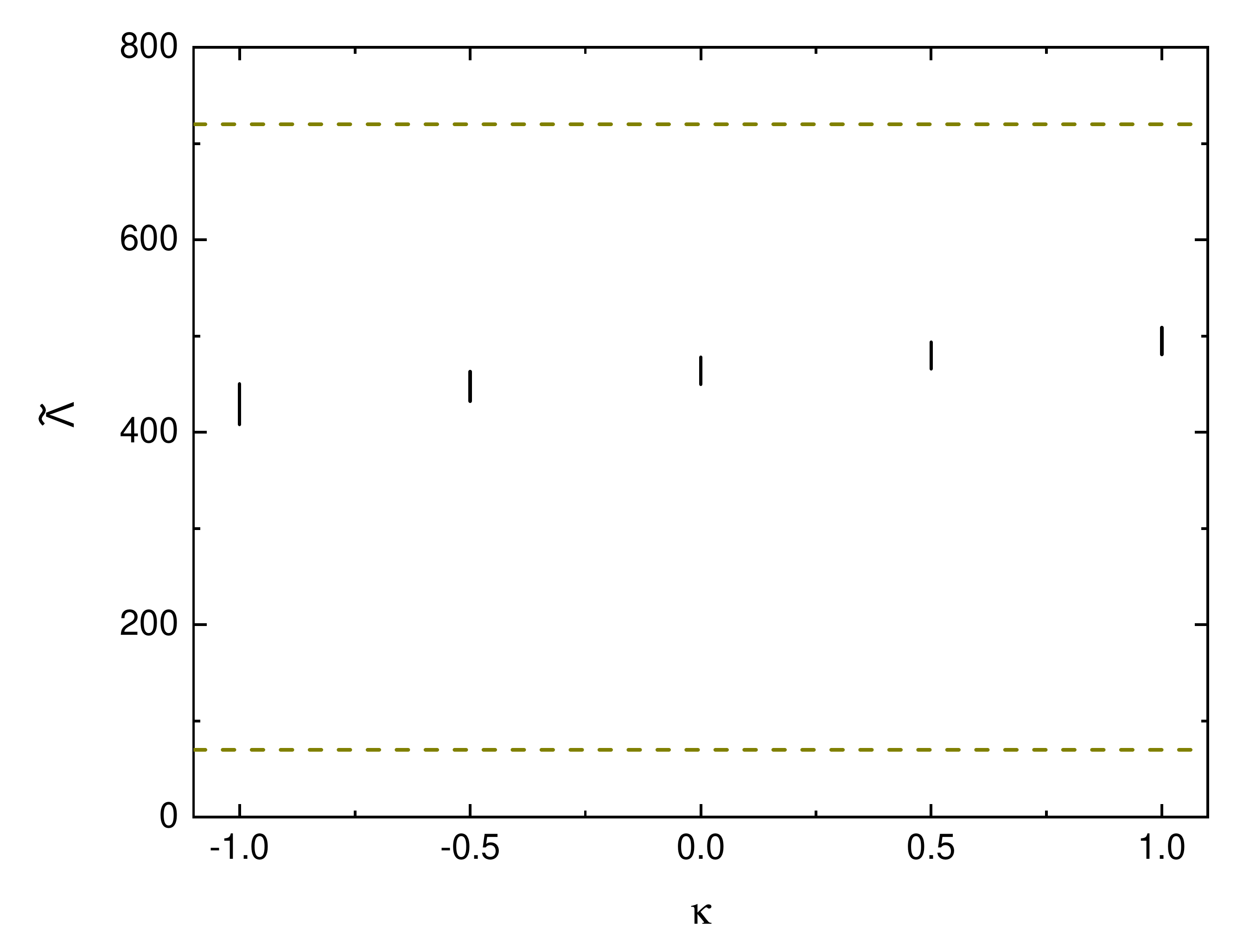} 
\caption{\label{fig_TD_tilde} ${\tilde\Lambda}$ for different values of of anisotropic parameter $k$. Dashed lines represent the range of ${\tilde\Lambda} = 300^{+420}_{-230}$ determined in the Ref.~\cite{abbott_2019}.}
\end{figure}

\section{Conclusions}\label{section4}

This work investigates the role of anisotropy on the fluid pulsation mode and tidal deformability of strange quark stars. This is realized through the numerical integration of the hydrostatic equilibrium, non-radial oscillations, and tidal deformability equations; which are modified to the inclusion of the anisotropy effects. To describe the fluid inside the star we assume the MIT bag model equation of state and to the anisotropy factor we employ the relation $\sigma=\kappa p_r(1-e^{-2\Psi})$.

Regarding the fluid pulsation modes, it is noted that $f$-mode changes considerably with the anisotropy, in contrast with the $p_1$-mode frequencies do not change much in the presence of anisotropy.

We also study the compatibility of dimensionless tidal deformability anisotropic strange stars with observational data reported by the LVC from the GW$170817$ event. In this scenario, we noted that the results reported in this article are within the set of observational data considered in this work. It is important to highlight that other anisotropy pressure profiles can be used in strange stars \cite{arbanil_malheiro_2016,aquino2022}, analyzing their dimensionless tidal deformability we can investigate the viability of the anisotropic profile and put some constraints using the same approach followed here. It should be noted that the deformability value increases with  $\kappa$ and decreases with $-\kappa$. This is in agreement with the study of polytropic stars investigated in \cite{arbanil_panotopoulos2022}. However, it is in discrepancy with studies reported in \cite{biswas2019} and \cite{das2022}, where the deformability profile is also investigated and how it changes with anisotropy $\sigma=\kappa(\rho+p_r)(\rho+3p_r)r^2 e^{2\Psi}/3$. From this, we can understand that the deformability increases or decreases with $\kappa$ depending on the type of anisotropic profile employed.

Additionally, it should be noted that in the literature we find works where the deformability parameter of strange stars is analyzed under different contexts. For example, in the reference \cite{lourenco2021} this parameter is analyzed considering quark matter in the color-flavor-locked (CFL) phase of color superconductivity, in \cite{li2021} this factor is investigated taking into account isospin effects in strange quark matter, and in the article \cite{xu2022} the tidal deformability are studied under the hypothesis that the quasiparticle model includes the non-perturbing characteristics of quantum chromodynamics in the low-density region. In the works in question, as well as in the present study, the light from the event GW1$70817$ is used to set limits to the study of strange stars under the backgrounds aforementioned.

Finally, it can be mentioned that the detectability of the oscillations modes is an important issue to be considered. This detectability is in a strong relationship with the parameters of the detectors, the most important being the sensitivity and frequency range. Moreover, it has to be considered that there are future planned upgrades for the actual operating LIGO-Virgo, in this case, the upgraded detector is called LIGO-Voyager \cite{Adhikari_2020}. In addition to this, it is well known that the scientific community has taken seriously the idea to build more technologically advanced gravitational wave detectors, we can mention the third-generation detectors: Einstein Telescope \cite{Punturo_2010}, Cosmic Explorer \cite{2019BAAS...51g..35R} and NEMO \cite{ackley_2020}. In this sense, the NEMO detector has a sensitivity of $10^{-24}\,[\rm Hz]^{-1}$. Therefore, its sensitivity is in the order of Cosmic Explorer and Einstein Telescope, but its technology primarily targets frequencies in the range of $1-4\,[\rm kHz]$, where is possible to observe the fundamental mode. As can be seen, with all the planned detectors, the observation of the oscillation modes of a compact star is a matter of time, and theoretical research in this direction is very important.


\begin{acknowledgments}
\noindent JDVA thanks Universidad Privada del Norte and Universidad Nacional Mayor de San Marcos for the financial support - RR Nº$\,005753$-$2021$-R$/$UNMSM under the project number B$21131781$. JMZP acknowledges financial support from the PCI program of the Brazilian agency Conselho Nacional de Desenvolvimento Científico e Tecnológico--CNPq.
\end{acknowledgments}


\appendix*

\section{Tidal deformability equations for the anisotropic case}

To derive the differential equations used to investigate the dimensionless tidal deformability for the anisotropic case, we start by considering the perturbed field equation:
\begin{equation}\label{perturbed_eq}
    \delta G^{\mu}_{\varphi}=8\pi\delta T^{\mu}_{\varphi},
\end{equation}
and, following Thorne and Campolattaro's work \cite{thorne1967}, we use the linear perturbation of the background metric tensor of the form
\begin{equation}
g^{(*)}_{\alpha\beta}=g_{\alpha\beta}+h_{\alpha\beta},
\end{equation}
with $g_{\alpha\beta}$ and $h_{\alpha\beta}$ standing the unperturbed metric tensor and the linearized perturbed metric, respectively. With these specializations, $h_{\alpha\beta}$ can be placed as \cite{thorne1967,regge1957}:
\begin{equation}
h_{\alpha\beta}={\rm diag}\left[He^{2\Phi},He^{2\Psi},r^2K,r^2K\sin^{2}\theta\right]Y_{\ell m},
\end{equation}
where $H=H(r)$ and $K=K(r)$ depend on the radial coordinate, and $Y_{\ell m}=Y_{\ell m}(\theta,\phi)$ is function of the angular coordinates. 

Expanding the fluid perturbation variables in terms of $Y_{\ell m}$, 
from the perturbed field equation \eqref{perturbed_eq} we found: 
\begin{widetext}
\begin{eqnarray}
&&\left[e^{-2\Psi}\left(K''-K'\Psi'-\frac{H'}{r}+\frac{3K'}{r}-\frac{H}{r^2}+\frac{2H}{r}\Psi'\right)-\frac{H\ell(\ell+1)}{2r^2}+\frac{K}{r^2}-\frac{K\ell(\ell+1)}{2r^2}\right]Y_{\ell m}=-8\pi\delta\rho,\label{G00(tr)}\\
&&\left[e^{-2\Psi}\left(K'\Phi'-2H\frac{\Phi'}{r}-\frac{H'}{r}+\frac{K'}{r}-\frac{H}{r^2}\right)+\frac{K}{r^2}+\frac{\ell(\ell+1)(H-K)}{2r^2}\right]Y_{\ell m}=8\pi\delta p_r,\label{G11(tr)}\\
&&\left[rHe^{2\Phi}\Psi'\Phi'-rHe^{2\Phi}\Phi'^2-rHe^{2\Phi}\Phi''+\frac{re^{2\Phi}H'\Psi'}{2}-\frac{3}{2}re^{2\Phi}H'\Phi'-\frac{re^{2\Phi}H''}{2}-\frac{re^{2\Phi}K'\Psi'}{2}\right.\nonumber\\
&&\left.+\frac{re^{2\Phi}K'\Phi'}{2}+\frac{re^{2\Phi}K''}{2}+He^{2\Phi}\Psi'-He^{2\Phi}\Phi'-H'e^{2\Phi}+K'e^{2\Phi}\right]\frac{e^{-2(\Psi+\Phi)}}{r}Y_{\ell m}=8\pi\delta p_t,\label{G22(tr)}\\
&&\left[\frac{H\Phi'}{r^2}+\frac{H'}{2r^2}-\frac{K'}{2r^2}\right]\partial_{\theta}Y_{\ell m}=0.\label{G12(tr)}
\end{eqnarray}
\end{widetext}
Substituting the Eq. \eqref{G12(tr)}, where $K'=2H\Phi'+H'$ and $K''=2H'\Phi'+2H\Phi''+H''$, in the subtraction of Eq. \eqref{G00(tr)}-Eq.\eqref{G11(tr)} and in the Eq. \eqref{G22(tr)}, we have, respectively:
\begin{widetext}
\begin{eqnarray}
&&-2Y_{\ell m}e^{-2\Psi}H\Psi'\Phi'-Y_{\ell m}e^{-2\Psi}H'\Psi'+2Y_{\ell m}e^{-2\Psi}H'\Phi'+2Y_{\ell m}e^{-2\Psi}H\Phi''+Y_{\ell m}e^{-2\Psi}H''+\frac{2H}{r}Y_{\ell m}e^{-2\Psi}\Psi'\nonumber\\
&&+\frac{6H}{r}Y_{\ell m}e^{-2\Psi}\Phi'+\frac{2H'}{r}Y_{\ell m}e^{-2\Psi}-\frac{H\ell(\ell+1)Y_{\ell m}}{r^2}-2Y_{\ell m}e^{-2\Psi}H\Phi'^2-Y_{\ell m}e^{-2\Psi}\Phi'H'=-8\pi\left(\delta\rho+\delta p_r\right),\label{delta_pr}\\
&&\frac{H}{r}e^{-2\Psi}Y_{\ell m}\left(\Psi'+\Phi'\right)=8\pi\delta p_t.\label{delta_pt}
\end{eqnarray}
\end{widetext}

For the perturbation of the radial pressure $p_r=p_r(p_t,\Psi)$, we have
\begin{equation}
\delta p_r=\frac{\partial p_r}{\partial p_t}\delta p_t,
\end{equation}
where it is considered $\delta\Psi=0$. In addition, $\delta\rho$ is defined by considering the equation of state $\rho=\rho(p_r)$. In this way, replacing equations \eqref{delta_pr} and \eqref{delta_pt} we obtain:
\begin{equation}\label{EDO_H}
H''+C_0H'+C_1H=0,
\end{equation}
where the functions $C_0=C_0(r)$ and $C_1=C_1(r)$ are calculated
as a function of the background quantities as follows
\begin{eqnarray}
&&C_0=\frac{2m}{r^2}e^{2\Psi}+4\pi e^{2\Psi}\left(p_r-\rho\right)r+\frac{2}{r},\label{C_0}\\
&&C_1=4\pi e^{2\Psi}\left[4\rho+4p_r+4p_t+\frac{p_r+\rho}{Ac_s^2}\left(c_s^2+1\right)\right]\nonumber\\
&&-\frac{\ell (\ell +1)}{r^2}e^{2\Psi}-4\Phi'^2,\label{C_1}
\end{eqnarray}
with $c_s^2=\frac{dp_r}{d\rho}$ and $A=\frac{dp_t}{dp_r}$.

\end{document}